\documentclass[useAMS,usenatbib]{mn2e}
\usepackage[dvips]{color,graphicx}
\newcommand{\bi}{\bfseries\itshape} 

\def\aj{AJ}             	
\def\apj{ApJ}           	
\def\aap{A\&A}          	
\def\mnras{MNRAS}       	

\def\afe{\rm[\alpha/Fe]}
\def\ofe{\rm[O/Fe]}
\def\feh{\rm[Fe/H]}

\title[Natures of a clump-origin bulge]{Natures of a clump-origin bulge: a pseudobulge-like but old metal-rich bulge}
\author[S. Inoue \& T. R. Saitoh]{Shigeki
Inoue$^{1,2}$\thanks{E-mail:inoue@astr.tohoku.ac.jp} \& Takayuki R. Saitoh $^{3}$
\\
$^{1}$Astronomical Institute, Tohoku University, Sendai 980-8578, Japan\\
$^{2}$Mullard Space Science Laboratory, University College London, Holmbury St. Mary, Dorking, Surrey, RH5 6NT\\
$^{3}$Interactive Research Center of Science, Tokyo Institute of
Technology, 2--12--1 Ookayama, Meguro, Tokyo 152--8551, Japan}

\begin{document}

\date{2011 June 16}

\pagerange{\pageref{firstpage}--\pageref{lastpage}} \pubyear{2002}

\maketitle

\label{firstpage}

\begin{abstract}
Bulges in spiral galaxies have been supposed to be classified into two types: classical bulges or pseudobulges. Classical bulges are thought to form by galactic merger with bursty star formation, whereas pseudobulges are suggested to form by secular evolution due to spiral arms and a barred structure funneling gas into the galactic centre. \cite{n:98,n:99} suggested another bulge formation scenario, `clump-origin bulge'. He demonstrated using a numerical simulation that a galactic disc suffers dynamical instability to form clumpy structures in the early stage of disc formation since the premature disc is expected to be highly gas-rich, then the clumps are sucked into the galactic centre by dynamical friction and merge into a single bulge at the centre. This bulge formation scenario, which is expected to happen only at the high-redshift, is different from the galactic merger and the secular evolution. Therefore, clump-origin bulges may have their own unique properties. We perform a high-resolution $N$-body/smoothed particle hydrodynamics (SPH) simulation for the formation of the clump-origin bulge in an isolated galaxy model and study dynamical and chemical properties of the clump-origin bulge. We find that the clump-origin bulge resembles pseudobulges in dynamical properties, a nearly exponential surface density profile, a barred boxy shape and a significant rotation. We also find that this bulge consists of old and metal-rich stars, displaying resemblance to classical bulges. These natures, old metal-rich population but pseudobulge-like structures, mean that the clump-origin bulge can not be simply classified into classical bulges nor pseudobulges. From these results, we discuss similarities of the clump-origin bulge to the Milky Way bulge. Combined with a result of \citet{ebe:08}, this pseudobulge-like clump-origin bulge could be inferred to form in clump clusters with a relatively low surface density.

\end{abstract}


\section{Introduction}
\citet{kk:04} has suggested that bulges in spiral galaxies can be classified into \textit{classical bulges} or \textit{pseudobulges}. Classical bulges are thought to form through violent dynamical relaxation by galactic merger like elliptical galaxies \citep[e.g.][]{nn:03,kk:04,nt:06,rkb:10,ho:10}. In other words, both of classical bulges and ellipticals could be merger remnants of galaxies, therefore supposed to be structurally identical. Indeed, their surface density profiles follow the $R^{1/4}$ law \citep{v:48}, their shapes are rounder than pseudobulges and they have little net rotation. Since the galactic merger is expected to consume most of gas in bursty star formation, the systems evolve passively and tend to be a single population system of old stars. The rapid star formation leads the enhancement of $\alpha$-elements and makes the bulge metal-rich \citep[][and references therein]{kk:04}. It is well known that classical bulges and ellipticals compose an identical fundamental plane and luminosity-metallicity relation \citep[e.g.][]{ki:83,k:85,jma:96,bbf:97}

Pseudobulges are discussed to form through secular evolution caused by non-axisymmetric structures in a galactic disc, such as spiral arms and a barred structure \citep[e.g.][]{cs:81,pn:90,ce:93,bf:02,dcm:04,a:05}. These non-axisymmetric structures rearrange angular momentum of the gas in a disc, funnel the gas into the galactic centre, which is presumed to arouse slow but long-lasting star formation resulting in a pseudobulge formation. Therefore, pseudobulges display a wide variety in their stellar population. As \citet{kk:04} has speculated that pseudobulges retain a memory of their discy origin, observations indeed show disc-like natures of pseudobulges, such as an exponential surface density profile, a significant rotation, a flatter shape than those of classical bulges and frequent existence of nuclear structures (e.g. bar, spiral and ring).

However, some bulges can not be simply classified into the two types above. The bulge of the Milky Way (MW) would be a typical example. Some observations have proposed a nearly exponential surface density profile, an oblate peanut shape (X-shape) and a significant rotation in the MW bulge, indicating the pseudobulge signatures (e.g. \citealt[][and references therein]{kdf:91,k:92,d:95,wgf:97,kk:04}; \citealt{ba:10,nug:10,mz:10,szw:11,nu:11,grz:11}). The MW bulge also has characteristic properties of classical bulges. It has long been known that the MW bulge consists of old stars of which ages are suggested to be approximately 11 -- 13 Gyr although uncertainty is still large (e.g. \citealt[][and references therein]{t:88,wgf:97,kk:04}; \citealt{s:06,b:10,b:11}). Additionally, rapid star formation episode were also suggested by metal-rich nature and an over-abundance of $\alpha$-elements in the MW bulge (e.g. \citealt[][and references therein]{wgf:97,kk:04}; \citealt{fws:03,no:06,z:06,zhl:08,rov:07,fmr:07,m:08,r:10,b:10,b:11,jrf:11,h:11}). Thus, these observations seem to indicate that the MW bulge can not be simply classified into classical nor pseudobulges. In addition, recent observations argued that some bulges, which were classified into pseudobulges, display exceptionally inactive star formation like classical bulges \citep{df:07,fdf:09,fd:10}. These inactive pseudobulges are also unclassifiable bulges.

\cite{n:98,n:99} has proposed another bulge formation scenario: \textit{`clump-origin bulge'}. He demonstrated with numerical simulations that since galactic discs are expected to be highly gas-rich in early stage of the disc formation, clumpy structures form due to instability of the gas in the gaseous disc, which could also explain some clumpy galaxies observed in the high-redshift Universe \cite[e.g.][]{n:98,n:99,isg:04,isw:04,bee:07,atm:09,abj:10,cdb:10,cdm:11}. These galaxies are referred to as clump clusters (chain galaxies) in the face-on (edge-on) view \citep[e.g.][]{bar:96,eeh:04,eem:09,gnj:11}. \cite{n:98,n:99} suggested that these clumpy stellar structures fall into the galactic centre by dynamical friction and merge into a single bulge at the galactic centre, being clump-origin bulge.

Clump-origin bulges form through `\textit{mergers of the clumps}' in a galactic disc, neither the galactic mergers nor the secular evolution. Therefore, properties of clump-origin bulges could be different from those of the conventional ones. Since the simulations of \cite{n:98,n:99} have been performed with a low resolution, he could not discuss the detailed properties of the clump-origin bulge (Noguchi, private communication). A few following studies discussed properties of the clump-origin bulge. \citet{ebe:08} argued with numerical simulations that the clump-origin bulges should be classified into classical bulges, since the bulges in their simulations have a density profile near to the $R^{1/4}$ law, a thick shape, and a slow rotation in stellar component although the clumps are fast rotating before the bulge formation. They suggested that initial \textit{spin} angular momentum of the clumps is lost to disc and halo when the clump-origin bulge forms. Recently \citet{cdm:11} studied dynamical state of gas in clumps migrating toward the galactic centre and demonstrated with both analytical approach and cosmological simulations that the gas component in most of clumps are highly rotation supported. Moreover, they discussed from their simulation results that more massive clumps are more highly rotation supported in the gas component and mentioned that clumps which have experienced a merger with other clumps tend to be highly rotationally supported ones (see their \S4.2). These results of \citet{cdm:11} imply that the clump merger spins up the gas in the remnant clump. Since the clump-origin bulge is the final remnant of the clump mergers, we expect that the clump-origin bulges might be rotating fast like pseudobulges. However, they did not discuss the final structure of the bulge or dynamical state of the stellar component.

We perform a similar (but more sophisticated and much higher resolution) numerical simulation to \cite{n:98,n:99} using an isolated galaxy model and study the naive natures of clump-origin bulges in details. We reconsider the rotation of the clump-origin bulge mentioned above, finally discuss similarity of the clump-origin bulge to the MW bulge. We describe our simulation settings in \S2 and the results of the bulge properties in \S3, including dynamical state, stellar age, metallicity and star formation activity. We present discussion in \S4 and summary of our results in \S5

\section{Simulation}
The detailed description of our computing schemes and initial conditions has been presented in \citet{is:11}. We describe it briefly here.

We use an $N$-body/SPH code, {\tt ASURA} \citep{sdk:08,sdk:09}. Gas dynamics is handled by the standard SPH scheme \citep{m:92,s:10}. We here adopt an individual time-step scheme, the time-step limiter of \citet{sm:09} to solve the evolution of the gas in the shocked regions correctly and the FAST method \citep{sm:10} which accelerates the simulations of self-gravitating gas by integrating the gravitational and hydrodynamical parts with different time-steps. Gravity is solved by a parallel tree with GRAPE (GRAvity PipE) method \citep{m:04}. We employ a barycentre approximation and the tolerance parameter of 0.5.

Cooling function of gas is assumed to be an optically thin radiative cooling, depending on metallicity and covers a wide temperature range of $10-10^8$ K \citep{wps:09}. Feedback from a uniform far-ultraviolet radiation and type-II supernovae (SNe) are also taken into account. A gas particle spawns a stellar particle of which mass is set to 1/3 of the mass of the original gas particle under the criteria: 1) $\rho_{\rm gas}>{\rm100~atm~cm^{-3}}$, 2) temperature $T<100$ K, 3) $\nabla\cdot${\bi v}$<0$ and 4) there is no SN around the particle. The local star formation rate is assumed to follow the Schmidt law \citep{s:59}. The non-dimensional star formation efficiency is set to $C_*=0.033$. Note that the global properties of star formation are fairly insensitive to the adopted value of $C_*$ when one adopts a high mass resolution ($\sim10^3M_\odot$) and a high threshold density for star formation such as $\rho_{\rm gas}>{\rm100~atm~cm^{-3}}$ \citep{sdk:08}. We assume the Salpeter initial mass function \citep{s:55} ranging from $0.1~{\rm M_\odot}$ to $100~{\rm M_\odot}$ on a stellar particle. We assume that stars heavier than 8 M$_\odot$ cause type-II SNe and thermally heat up ambient gas (see also \S3.3).

Our initial condition follows the spherical model of \citet{kmw:06,kmw:07} that was used to study the formation of disc galaxies in an isolated environment. We assume an equilibrium system with the Navarro-Frenk-White profile \citep{nfw:97} with a virial mass $M_{\rm vir} = 5.0\times10^{11}~{\rm M_\odot}$, a virial radius $r_{\rm vir} = 1.67\times10^2$ kpc and a concentration parameter, $c=6.0$. Baryon mass fraction of the system is set to $0.06$. Initially all of baryon is primordial (zero-metal) gas having a virial temperature. The density profiles of dark matter (DM) and gas components are initially the same. The gas sphere has a rotation following a specific angular momentum distribution of $j\propto r$ \citep{bdk:01} normalized by a spin parameter of $\lambda=0.1$. The DM halo is represented by $10^7$ particles with a softening length of $\epsilon_{\rm DM} = 8~{\rm pc}$, whereas the gas in the halo is expressed by $5.0\times10^6$ particles with $\epsilon_{\rm gas} = 2~{\rm pc}$. The masses of DM and gas particles are $4.7\times10^4M_\odot$ and $6.0\times10^3M_\odot$, respectively.

Although we solve the wide temperature range of the inter-stellar matter (ISM), $T=10 - 10^8~{\rm K}$, we adopt neither temperature nor pressure floor. This is because we think these artificial suppression techniques of the low Jeans mass objects are unnecessary. It is widely believed that insufficient mass resolution may lead to unphysical fragmentation in the ISM. To avoid this \textit{'unphysical fragmentation'}, one may adopt temperature/pressure floors which can maintain the condition of $M_{\rm Jeans}>N_{\rm nb}\times m_{\rm gas}$, where $M_{\rm Jeans}$ is the Jeans mass, $N_{nb}$ is the number of neighbouring SPH particles and $m_{\rm gas}$ is the mass of a gas particle. Since the original claim of the unphysical fragmentation was done based on their simulations during non-linear evolutions \citep{bb:97}, it is hard to clarify what was the true origin of the unphysical fragments. This problem has been first pointed out by \citet{bb:97}. Since their evaluation was based on their simulation during non-linear evolutions, it is difficult to clarify what was the actual origin of the unphysical fragments.

\citet{hgw:06} showed with well-conceived numerical tests for this problem that the growth of a linear-regime perturbation in a self-gravitating fluid is just slower than that of the analytical solution when the mass resolution is insufficient to express the local Jeans mass ($M_{\rm Jeans}>N_{/rm nb}\times m_{\rm gas}$). They did not find any unphysical fragmentation in simulations with insufficient mass resolution. Therefore, there is no need to worry about the unphysical fragmentation in simulations with insufficient mass resolution without temperature/pressure floor. Hence, the result that the disk holds a number of small clumps especially in early phase in our simulation is not a numerical artifact (see, \S 3).

The effect of the insufficient mass resolution in our simulation would appear as the suppression of growth rates of clump seeds. However, this is not crucial for our results because (1) the dynamical time of seeds we may drop in our simulation is short because they are compact (2) the final mass of the clumps are much larger than the resolution limit.

\section{Results}
\begin{figure*}
  \begin{minipage}{170mm}
    \includegraphics[width=170mm]{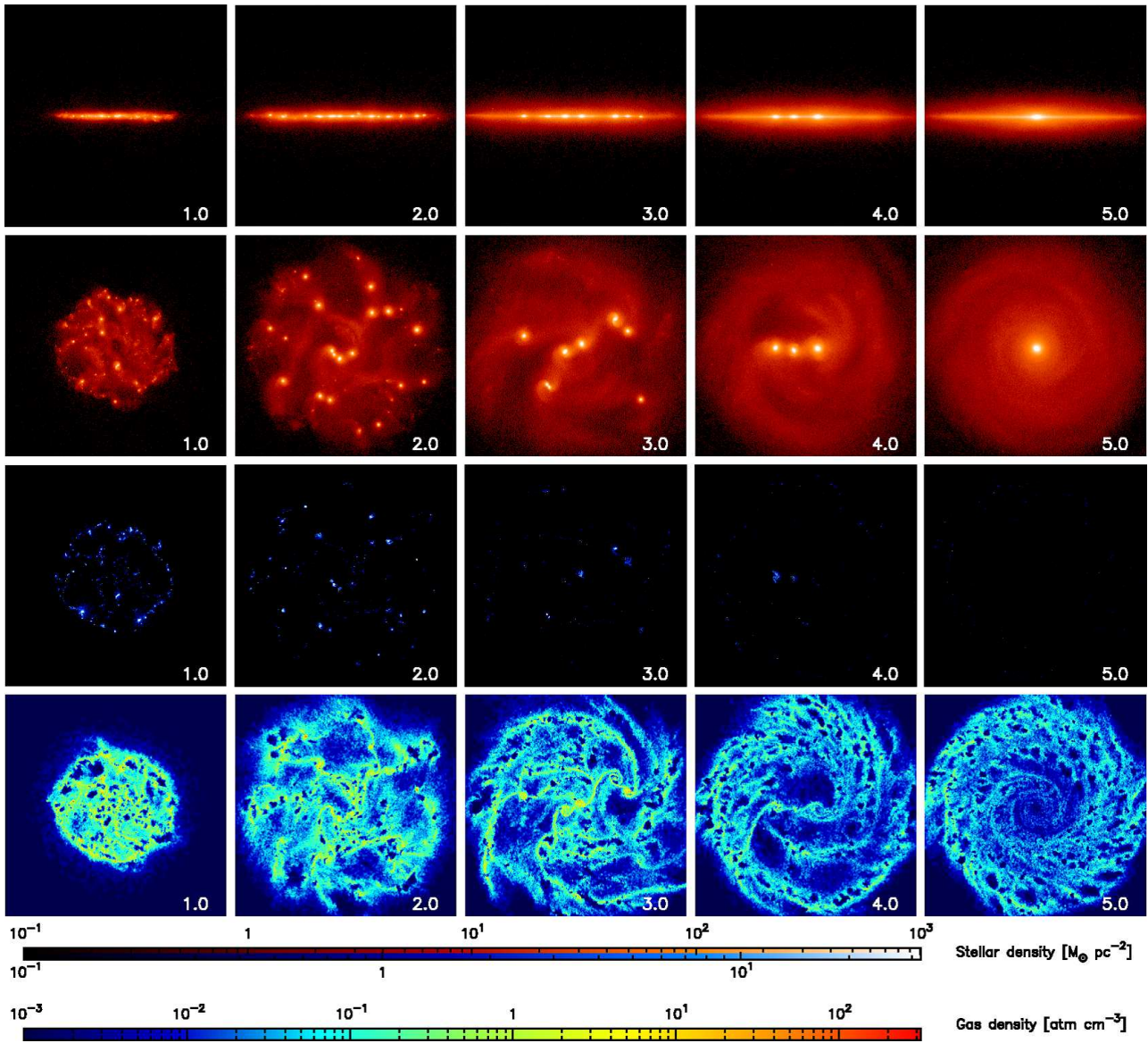}
   \caption{Stellar and gas density maps in the central $60\times60$ kpc region. The first and second rows indicate the surface density of all stars from the edge-on and face-on views, respectively. The third row plots the density of stars younger than 40 Myr as star forming region. The fourth row shows the volume density of gas in a slice of the disc plane. Time in unit of Gyr is indicated on right bottom in each panel.}
   \label{snap}
  \end{minipage}
\end{figure*}

We run the simulation for 6 Gyr and display snapshots of stellar and gas distributions and star forming region in Fig. \ref{snap}. The simulated galaxy settles into a stable state after 5 Gyr. 

At $t\hspace{0.3em}\raisebox{0.4ex}{$<$}\hspace{-0.75em}\raisebox{-.7ex}{$\sim$}\hspace{0.3em}2~{\rm Gyr}$, many clumps form in the disc onto which the surrounding gas is falling. Active star formation can be seen only in the clumps, whereas little star formation is found in the inter-clump (disc) region (the third row). This clumpy phase is seen in previous numerical studies and suggested to explain the clump clusters and chain galaxies observed in the high-redshift Universe \citep[e.g.][]{n:98,n:99,isg:04,isw:04,bee:07,atm:09,cdb:10,cdm:11,abj:10}. In our simulation model, since the system is lacking continuous gas supply supposed in the cosmological context, the gas accretion ceases at $t\sim2~{\rm Gyr}$. After the cessation of gas supply, only a small number of tiny clumps form in the disc, which are fragile to tidal force and these disrupted clumps build up a stellar disc. The other massive clumps can survive, merge with one another and grow to more massive ones. In this phase, a central density slope of the DM halo is turned cored by the reaction of the dynamical friction on the clumps \citep{is:11}. The clumps are sucked into the galactic centre by dynamical friction and finally merge into a single \textit{`clump-origin'} bulge as first suggested by \citet{n:98,n:99}. At the final state ($t=6$ Gyr), 80 per cent of the initial gas had converted into stars; the total stellar and gas masses are $2.5\times10^{10}~{\rm M_\odot}$ and $5.0\times10^{9}~{\rm M_\odot}$, respectively. There is little gas in the halo region. The central density cusp of the DM halo is revived due to the gravitational potential of the massive bulge \citep{is:11}.

\citet{n:96} has first suggested a thick disc formation scenario by dynamical heating on a disc by clumps, and \citet{ee:05} observationally discussed the existence of a thick disc in chain galaxies. \citet{bem:09} demonstrated with numerical simulations that a disc forming in clump clusters is a thick disc, whereas a thin disc forms via gas accretion at the latter epoch than that of the thick disc formation. Since our simulation model does not take such a late stage accretion into account, we indeed find in our simulation that the disc is dominated by the thick disc component with a scale height of $1.41~{\rm kpc}$ and a scale radius of $9.85~{\rm kpc}$, while a thin disc hardly forms. Therefore, in this paper, we consider the disc in our simulation to be a single-component thick disc which forms via the clumps. However, it does not mean to reject other possible scenarios of thick disc formation, such as dissipative collapse \citep[e.g.][]{bth:92}, thin disc heating \citep[e.g.][]{qhf:93}, accretion of dwarf galaxies \citep[e.g.][]{ans:03}, multiple gas-rich minor merger \citep{bkg:04,bgm:05} and radial migration \citep[e.g.][]{sb:02}. We will report all the details of the thick disc formation and kinematics in our simulation in a forthcoming paper (Inoue \& Saitoh, in prep).

\begin{figure}
  \includegraphics[width=85mm]{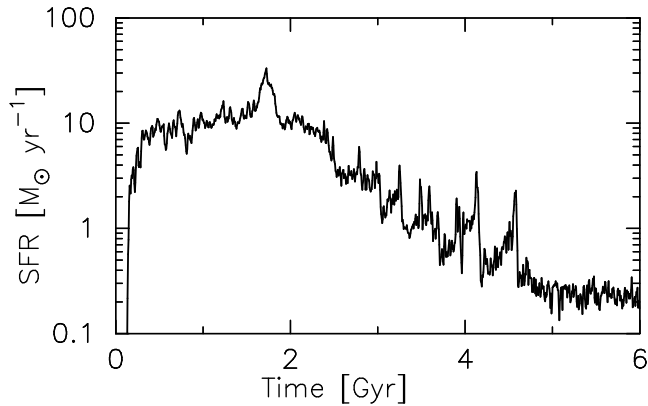}
  \caption{Star formation rate (SFR) as a function of time. The entire region of the galaxy is considered.}
  \label{sfr}
\end{figure}

In Fig. \ref{sfr}, we show the star formation history of the entire galaxy. Since the gas is continuously falling onto the galaxy ($t\hspace{0.3em}\raisebox{0.4ex}{$<$}\hspace{-0.75em}\raisebox{-.7ex}{$\sim$}\hspace{0.3em}2~{\rm Gyr}$), the system keeps star formation rate of $\sim10~{\rm M_\odot~yr^{-1}}$. After the gas infall phase, the galaxy asymptotically settles to the low star formation state with $\sim$ 1 - 0.1 ${\rm M_\odot~yr^{-1}}$. When the stellar clumps coalesce with one another, the star formation activity is temporarily activated in the clusters, resulting in appearance of spiky features in the late history of star formation in Fig. \ref{sfr}. 

\subsection{Dynamical state}
\subsubsection{Density profile}
In order to identify the bulge type, i.e. classical or pseudobulges, the S\'ersic profile fitting of a surface density profile of bulge stars is frequently used:
\begin{equation}
\Sigma(R) = \Sigma_0\exp\left[-\left(\frac{R}{R_0}\right)^\frac{1}{n}\right],
\end{equation}
where $\Sigma_0$ and $R_0$ are the central surface density and the scale radius of the bulge, respectively, and $n$ is the S\'ersic index. If $n=1$, the profile is consistent with an exponential profile implying a pseudobulge. If $n=4$, the profile follows the de Vaucouleurs' $R^{1/4}$ law \citep{v:48} implying a classical bulge. The boundary is empirically set to $n\simeq2$ with little overlap \citep[e.g.][]{kk:04,df:07,fd:08}. 

\begin{figure}
  \includegraphics[width=90mm]{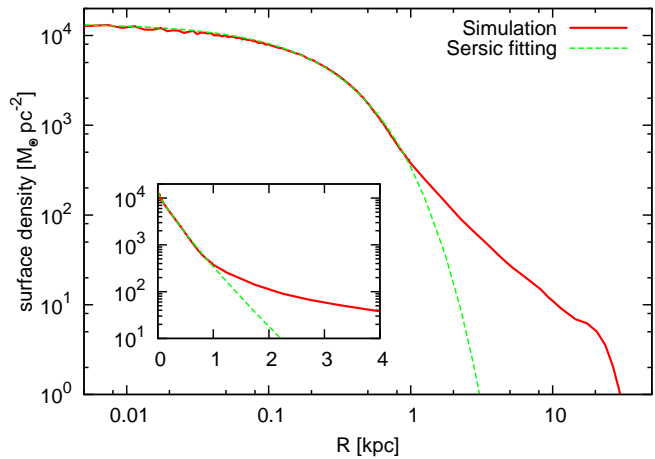}
  \caption{Stellar surface density profile from the face-on view at the final state of the simulation ($t=6~{\rm Gyr}$). Minimum $\chi^2$ fitting by the S\'ersic profile with the index, $n=1.18$, is also plotted. The centre of the system is defined to be the densest point in the stellar density distribution. The insert diagram illustrates the same density profiles with the linear abscissa.}
  \label{Sersic}
\end{figure}
In Fig. \ref{Sersic}, we plot the azimathally averaged surface density of stars at the final state and the result of the S\'ersic profile fitting. The fitting by minimum $\chi^2$ is given by $\Sigma_0 = 1.37\times10^4~{\rm M_\odot/pc^2}$, $R_0 = 0.21~{\rm kpc}$, and the S\'ersic index, $n = 1.18$. This value of the index, $n$, means the bulge in our simulation has a nearly exponential surface density profile and indicates a pseudobulge-like density structure. Contrary to this, \citet{ebe:08} has indicated with numerical simulations that their clump-origin bulges have a S\'ersic index larger than 2 and the morphology of classical bulges. We will discuss this contradiction in \S4.4.

The S\'ersic profile can not fit the simulated profile outside $R\hspace{0.3em}\raisebox{0.4ex}{$>$}\hspace{-0.75em}\raisebox{-.7ex}{$\sim$}\hspace{0.3em}1~{\rm kpc}$. We find the transition radius between the disc and the bulge components at $R\sim1~{\rm kpc}$ on the surface density. Thus, we define the bulge region to be inside $R=1.0~{\rm kpc}$. The bulge has a mass of $4.5\times10^9$ M$_\odot$ and a half-mass radius of $450~{\rm pc}$. Ratio of the bulge mass to the total stellar mass, $B/T$, is 0.18\footnote{For example, the MW has a value of $B/T=0.19\pm0.02$ \citep{kdf:91,d:95}.}. However, it must be kept in mind that the galaxy in our simulation does not have a thin disc. Since a thin disc accounts for not a small fraction of stellar mass of a real galaxy, the $B/T$ must be overestimated in our simulation. \citet{fd:08} has observed the mean value of $B/T$ to be $\langle B/T \rangle=0.16\pm0.05$ among pseudobulges and $\langle B/T \rangle=0.41\pm0.11$ among classical bulges. Therefore, the clump-origin bulge in our simulation is near to pseudobulges on the value of $B/T$.

It should be noted that the $B/T$ could be sensitive to simulation setup. We speculate that the $B/T$ would become smaller if the gas accretion onto the disc is slower, because wavelength at which instability first appears becomes smaller in a gas-poorer disc \citep{t:64,bt:08}, which allows to form smaller clumps preferentially. They tend to be disrupted by the galactic tide and/or lead to result in a smaller clump-origin bulge. On the other hand, if we assume a smaller spin parameter in the initial condition, the gas forms a smaller disc and the gas is more condensed in the disc than that in the current simulation \citep{fe:80,bw:87}. In such cases, the wavelength of the instability becomes larger, the larger clumps would form and result in the formation of a larger bulge. In the dense and small disc, since the timescale of the clump migration becomes much shorter, the bulge formation would be finished earlier than the current simulation.

\subsubsection{Shape}
\begin{figure}
  \includegraphics[width=80mm]{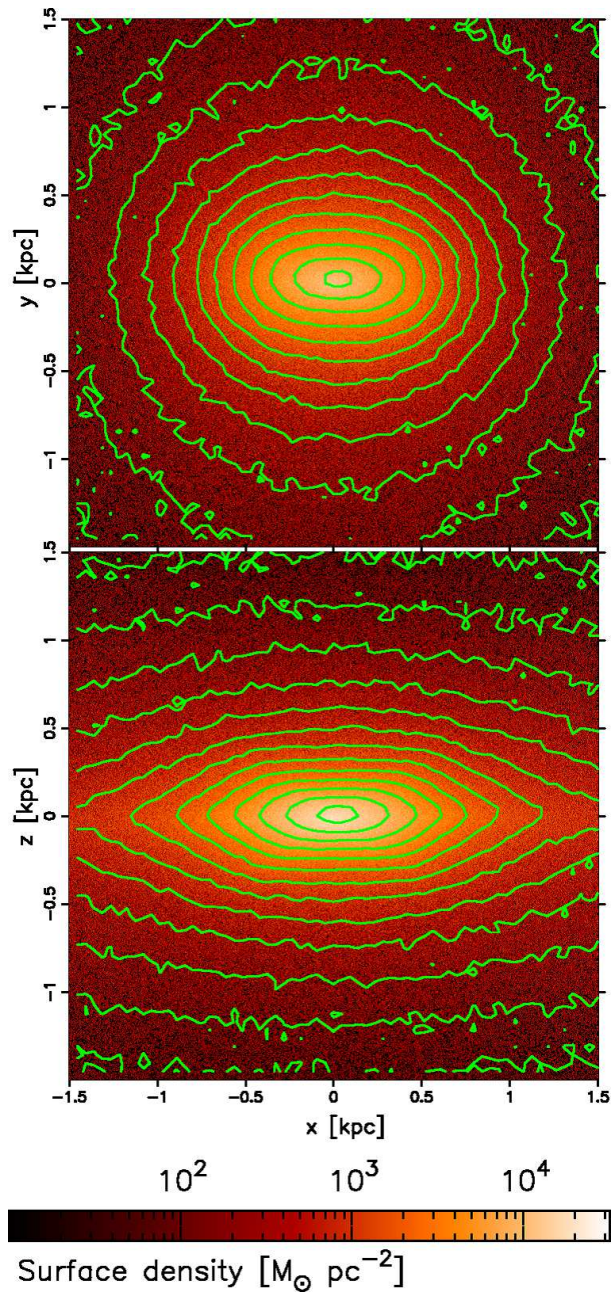}
  \caption{Stellar surface density maps in central $3\times3$ kpc region at the end-state of the simulation ($t=6~{\rm Gyr}$). The top panel is from the face-on view and the bottom one is from the edge-on view. The horizontal axes ($x$-axes) coincide with the major-axis of the bulge in both panels. The contour levels in the images are chosen to highlight the bulge shape.}
  \label{bulge_zoom}
\end{figure}

In Fig. \ref{bulge_zoom}, we show stellar surface density maps of the bulge at the final state in our simulation. As seen in the image from the face-on view (the top panel), the bulge has a bar-like elongated feature. This `nuclear bar' in a bulge is a common feature among pseudobulges \citep[e.g.][]{kk:04}. Ellipticity\footnote{In this paper, the ellipticity is defined to be $1-b/a$ of the isodensity contour at which the surface density is equal to $\Sigma_0/2$, where $a$ and $b$ are lengths of major- and minor-axes.} of the bulge measures 0.55 from the face-on view.

As seen from the edge-on view (the bottom panel), it clearly appears that this bulge is a boxy bulge, indicating a pseudobulge signature \citep[e.g.][]{kk:04,a:05}. Ellipticity of the bulge measures 0.67 from the edge-on view.

\subsubsection{Rotation}
\begin{figure}
  \includegraphics[width=77mm]{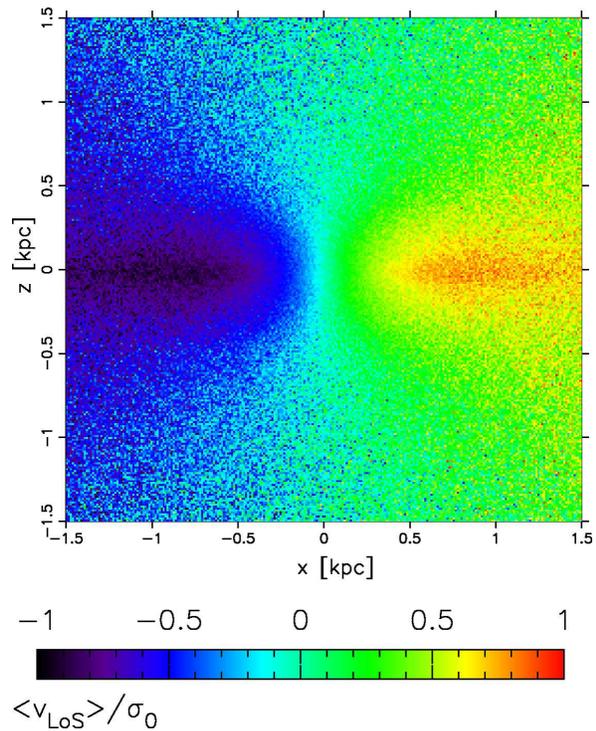}
  \caption{Mean LoS velocity map of the stellar component in the central $3\times3$ kpc region at the final state of the simulation ($t=6~{\rm Gyr}$) from the angle edge-on and perpendicular to the major-axis of the bulge. The value is normalized by the LoS velocity dispersion interior to the half mass radius of the bulge, $\sigma_0=81.0~{\rm km~s^{-1}}$.}
  \label{vel}
\end{figure}

Pseudobulges are generally observed to have a significant rotation (e.g. \citealt[][and references therein]{kk:04}; \citealt{wzb:11}). On the other hand, classical bulges are generally presumed not to rotate significantly although some classical bulges show a net rotation \citep{ki:82,c:07}.

In Fig. \ref{vel}, we illustrate a map of the mean line-of-sight (LoS) velocity of stars from the edge-on view (perpendicular to the major-axis) at the final state. The value is normalized by the LoS velocity dispersion interior to the half mass radius of the bulge ($450~{\rm pc}$), $\sigma_0=81.0~{\rm km~s^{-1}}$. This bulge shows a significant rotation. The value of $V_{\rm max}/\sigma_0\simeq0.9$ of this bulge means that the rotation (spin) is \textit{not} negligible but comparable to the velocity dispersion, where $V_{\rm max}$ is the maximum mean LoS velocity in the bulge. This is also a pseudobulge-like signature \citep[][and references therein]{kk:04}. The direction of the rotation coincides with the disc rotation.

\begin{figure}
  \includegraphics[width=90mm]{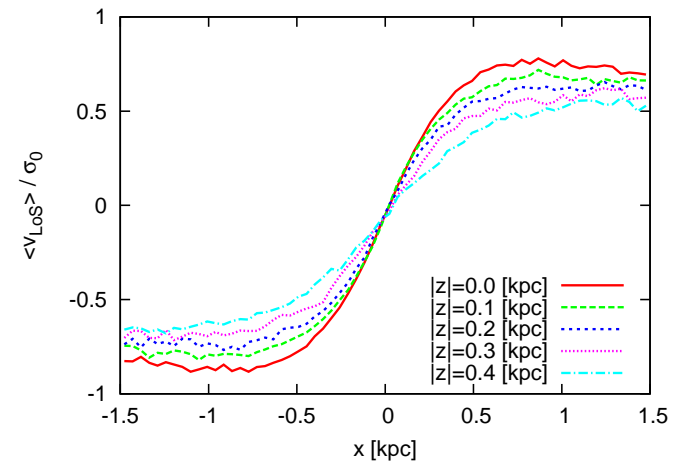}
  \caption{Rotation velocity curves at different heights above the disc plane, $z$. The abscissa coincides with the major axis of the bulge.}
  \label{cylindrical}
\end{figure}

Recently, observations of some pseudobulges have shown that rotational velocities do not depend on a height within the bulges \citep[][and references therein]{wzb:11}, including the MW bulge \citep{h:09}, i.e. `\textit{cylindrical rotation}'. However, regardless of our results that the clump-origin bulge follows the pseudobulge-like natures and the boxy morphology, we do \textit{not} find a cylindrical rotation in the clump-origin bulge simulated in this work. Fig. \ref{cylindrical} shows that the rotation velocity curves tend to be flatter at a higher latitude of the bulge. However, \citet{wzb:11} has observationally shown that boxy/peanut-shaped bulges do not necessarily involve a cylindrical rotation. Therefore, their result implies that the cylindrical rotation may not be a congenital nature of the boxy/peanut-shaped bulges.

Recently, \citet{smg:11} indicated with a numerical simulation that interaction with a barred structure in a disc can spin up a non-rotating bulge and the bulge acquires a boxy/peanut shape and a cylindrical rotation. If this mechanism is efficient, the cylindrical rotation can be built up after the bulge established and may not be a strong indicator for the bulge formation mechanism.

\subsection{Age}
As we noted in \S 1, classical bulges are generally an old structure because the classical bulges form by past galaxy merger events. Therefore, their stellar evolution tends to be passive after the last merger. As for pseudobulges, their star formation is generally still going on (although see also \S3.4) because non-axisymmetric structures continue to funnel gas into the central bulge. Hence, stellar age in a bulge is also another important indicator to identify the bulge type. 

\begin{figure}
  \includegraphics[width=90mm]{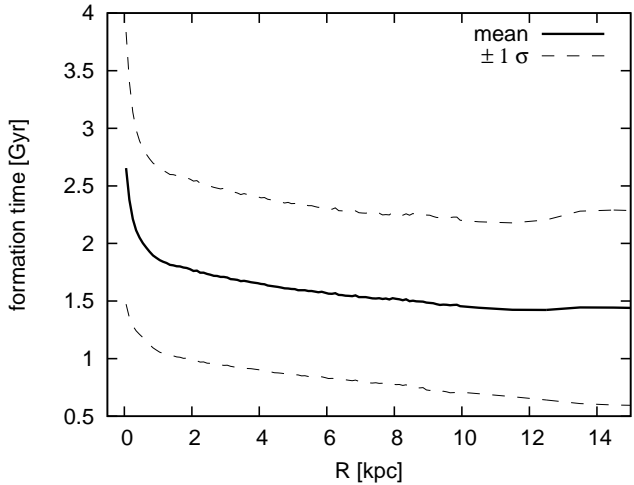}
  \caption{Averaged formation time of star particles as a function of the galactocentric distance. The values are azimathally averaged and a dispersion, $\sigma$, is calculated in each radial bin. This is a formation time of stars, not age. We define the start of the simulation to be the zero-point of the formation time.}
  \label{age}
\end{figure}

In Fig. \ref{age}, we plot the mean and dispersion of the formation time of stars in our simulation. As seen in the figure, the bulge ($R\hspace{0.3em}\raisebox{0.4ex}{$<$}\hspace{-0.75em}\raisebox{-.7ex}{$\sim$}\hspace{0.3em}1~{\rm kpc}$) is younger than the thick disc only by $\sim1~{\rm Gyr}$. It is because the clumps in the clump cluster retain star formation activity inside them (the third row of Fig. \ref{snap}) and migrate to the galactic centre while forming new stars. The formation time of disc stars are almost independent from the distance from the galactic centre.

However, the difference in the formation time between the bulge and the disc, $\sim1~{\rm Gyr}$, is so small in comparison with the age of the Universe that we could presume the bulge to form simultaneously with the disc. As mentioned above, the disc in our simulation is a thick disc, does not have a thin disc. Moreover, this bulge formation scenario is expected to happen only in the high-redshift Universe. Therefore, we could consider that this clump-origin bulge is an old structure, as old as the thick disc.

To be an old structure \textit{does} conflict with the morphological and dynamical similarities of this clump-origin bulge to pseudobulges mentioned in \S3.1. Therefore, we suggest that the clump-origin bulge dynamically resembles pseudobulges but consists of the old stars like classical bulges. Hence, the clump-origin bulge can \textit{not} be simply classified into classical bulges nor pseudobulges, but shows peculiar properties. However, we have to note that the bulge may be rejuvenated by the late accretion to form a thin disc.

\subsection{Metallicity}
Before discussing metallicity, we should describe here how we treat chemical evolution in our simulation. Our simulation code tracks only the total metal abundance, $Z$, and does not follow the evolution of each element. The initially primordial gas is contaminated with heavy elements by type-II SNe. We did not take type-Ia (or other) SNe or stellar wind from intermediate mass stars into account in our simulation.

\begin{figure}
  \includegraphics[width=90mm]{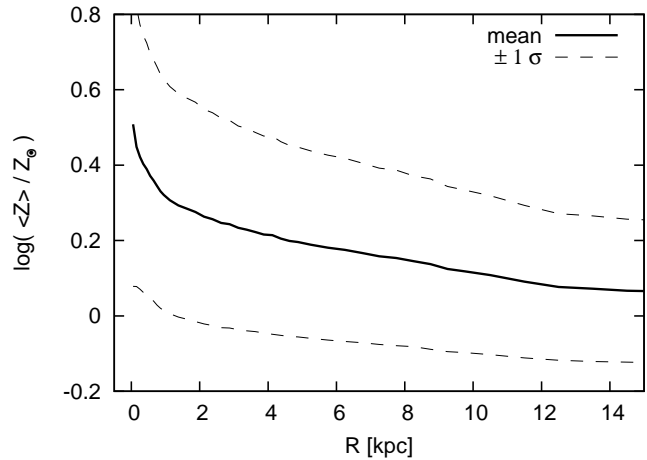}
  \caption{Radial metallicity distribution at $t=6~{\rm Gyr}$. The dashed lines indicate the range of $\pm1\sigma$ dispersion. The value of solar metallicity, $Z_\odot$, is set to 0.02.}
  \label{Z}
\end{figure}

We show a radial metallicity distribution in our simulation in Fig. \ref{Z}. Stars of the clump-origin bulge are more metal-rich by 0.2 dex than stars of the inner region of the thick disc. Such a high metal abundance in the bulge is a natural outcome of the formation of clumps which form from a collapsing gas cloud in the highly gas-rich disc and the violent star formation in the clump mergers.

Recent observations have shown that there is no or little chemical distinction between the MW bulge and the thick disc stars in $\afe$ vs. $\feh$ distributions despite exhibiting clear differences in their metallicity distribution functions \citep{m:08,b:09,bao:10,b:10,b:11,ama:10,g:11,h:11}. However, observation results of \citet{lhz:07} and \citet{fmr:07} indicated clear separations of $\afe$ between the thick disc and bulge stars in the MW. Observations of the chemical distinction between the MW bulge and the thick disc stars are still controversial because of only a handful number of reliable sample stars and lack of accuracy of distances and proper motions to distinguish the bulge and disc stars\footnote{\citet{h:11} argued that using different reference stars in calibration was the cause of the conflict.}. Our simulation predicts the higher metallicity in the bulge than the thick disc. More convincing observations are highly anticipated to test such a scenario. Note, however, that there are many ways to build up thick discs, as we noted in \S2. 

\citet{r:11} recently observed that the MW thick disc shows a small radial gradients in $\feh$ ratio, $+0.01\pm0.04~{\rm dex~kpc^{-1}}$. \citet{abw:06}, \citet{i:08} and \citet{bao:11} also observationally confirmed no or little radial $\feh$ gradient in the thick disc stars. Our result in Fig. \ref{Z} shows a small negative gradient, $-0.014~{\rm dex~kpc^{-1}}$, in metallicity, $Z$, from $R=2.0~{\rm kpc}$ to $15.0~{\rm kpc}$, which is not inconsistent with these observations in the MW within the error. 

\begin{figure}
  \includegraphics[width=90mm]{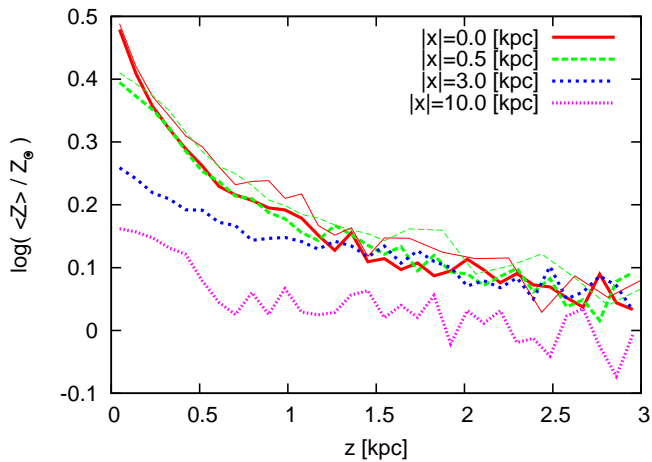}
  \caption{Vertical metallicity distribution at the bulge regions ($|x|$=$0.0~{\rm kpc}$ and $0.5~{\rm kpc}$) and disc regions ($|x|$=$3.0~{\rm kpc}$ and $10.0~{\rm kpc}$). Values are averaged along LoS. Thin lines (red solid and green long-dashed) indicate the metallicity distribution of the bulge stars only ($R<1~{\rm kpc}$).}
  \label{Z_vert}
\end{figure}

It has been indicated that there is a clear variation in metallicity along the minor axis of the MW bulge \citep{bh:74,t:88,ftb:90,tfw:90,zhl:08,jrf:11,grz:11}. \citet{zhl:08} observed the vertical gradient of the mean $\feh$ ratio to be $-0.6~{\rm dex~kpc^{-1}}$. We here show metallicity distributions along the vertical direction in our simulation in Fig. \ref{Z_vert}. This figure clearly indicates a vertical metallicity gradient along the minor axis in the clump-origin bulge, $z\hspace{0.3em}\raisebox{0.4ex}{$<$}\hspace{-0.75em}\raisebox{-.7ex}{$\sim$}\hspace{0.3em}1.0~{\rm kpc}$ of the solid red line. Contamination by the disc stars is negligible in the bulge region (the thin lines). In our result, the metallicity keeps increasing even in the central region of the clump-origin bulge although \citet{rov:07} has observed no evidence for the vertical gradients of $\feh$ and $\afe$ in the central region of the MW bulge, between $(l,b)=(1^\circ ,-4^\circ )$ and $(l,b)=(0^\circ ,-1^\circ )$ in the Galactic coordinate.

In the MW thick disc, \citet{r:11} observed that the vertical gradient of $\feh$ ratio is very small, $-0.09 \pm 0.05~{\rm dex~kpc^{-1}}$. In our simulation, the vertical gradient of $Z$ is small, $\sim -0.05~{\rm dex~kpc^{-1}}$ in the thick disc region. Additionally, we find that the vertical metallicity gradients are almost independent of the galactocentric distance. This is represented in Fig. \ref{Z_vert}. The blue dotted line (inner disc region) and the purple dot-dashed line (outer disc region) are almost parallel.

\subsection{Star formation activity}
Classical bulges are supposed to have used up their gas and/or have a central massive black hole to heat up and prevent gas from accreting \citep{onb:08}. Therefore, they are generally inactive in star formation. Pseudobulges are still forming stars actively, which is observationally verified by star forming nuclear rings nuclear spirals, bars and ovals, implying similarities to disc galaxies (e.g. \citealt[][and references therein]{kk:04}; \citealt{f:06,fd:10}). 

\begin{figure}
  \includegraphics[width=90mm]{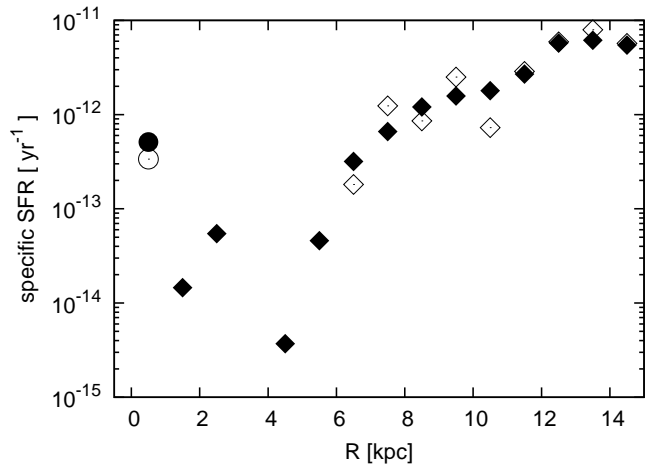}
  \caption{Specific star formation rate at the final state of the simulation ($t=6~{\rm Gyr}$). Filled and open symbols indicate values averaged in the last 0.5 Gyr and 40 Myr, respectively. Circles and diamonds indicate the values of the bulge and disc region, respectively. The open symbols are missing data points at inner disc region, $1~{\rm kpc}<R<6~{\rm kpc}$, since no star particle was born in this radial range in the last 40 Myr.}
  \label{SSFR}
\end{figure}

In Fig. \ref{SSFR}, we show the specific star formation rate\footnote{The specific star formation rate is defined in this paper as a star formation rate in a radial bin divided by the mass of the bin.} in the bulge and the thick disc of our simulation. The clump-origin bulge is little forming stars at the final state. This bulge has very quiet and weak star formation activity. In the thick disc, the specific star formation rate becomes larger as $R$ increases. The innermost region of the disc is less active than the bulge region. There is little difference between the values averaged in the last 40 Myr (the open symbols) and 0.5 Gyr (filled symbols). Thus, the inactive star formation is not transient. This halted star formation in the bulge is a typical feature of classical bulges.

Some bulges classified into pseudobulges are discovered to show inactive star formation like classical bulges \citep{df:07,fdf:09,fd:10}. These bulges have a small value of the S\'ersic index, $n\hspace{0.3em}\raisebox{0.4ex}{$<$}\hspace{-0.75em}\raisebox{-.7ex}{$\sim$}\hspace{0.3em}2$, despite the fact that their star formation activity is as weak as classical bulges. These are called `\textit{inactive pseudobulges}' \citep{fdf:09,fd:10}. Although the bulge in our simulation is less active than the observed inactive pseudobulges by $\sim1~{\rm dex}$ in the specific star formation rate ($\sim 10^{-11}~{\rm yr^{-1}}$ in the observations), the clump-origin bulge in our simulation resembles to the inactive pseudobulges: the nearly exponential surface density and the inactive star formation.

We have to note that it could not be fair to directly compare the bulge in our simulation with the inactive pseudobulges in observations. The inactivity of these observed pseudobulges may be due to feedback from an active galactic nucleus, while our simulation does not take it into account. Moreover, the simulated galaxy is lacking the later gas accretion which would play an important role for the thin disc formation. The star formation activity in the clump-origin bulge would be affected by the gas accretion: a dynamically cold thin disc forms non-axisymmetric structures such as spiral arms and a galactic bar, which can supply the bulge with gas more or less and would activate the star formation in the bulge.

\section{Discussion}

\subsection{Validity of the model settings and comparison with observations}
We set the larger spin parameter than the mean value of cosmological simulations, $\lambda\simeq0.035$, \citep[e.g.][]{bdk:01,kds:11} although observed values among high-redshift galaxies are still controversial \citep[e.g.][]{bo:07,du:11}. As we mentioned in \S3.1.1, the large spin parameter would lead the formation of a large disc, small clumps and a small bulge. \citet{ee:05} observed that diameters of galactic discs range 13 -- 29 ${\rm kpc}$ among 10 clump cluster samples of which masses are 0.5 -- 25 $\times10^{10}M_\odot$. The numbers of clumps in the sample are 5 -- 14 per galaxy and the averaged clump masses in the clump cluster range 0.21 -- 1.36 $\times10^9M_\odot$. The disc in our simulation is larger than their observational result by a factor of 2 or 3 and the surface density of inter-clump region is lower than the observed values \citep[see also][]{eem:09}. Therefore, our simulation may \textit{not} be a well-reproduced model of the high-redshift galaxies in the current observations on the points of the disc size and the surface density, although our simulation does not seem inconsistent with the observation of \citet{ee:05} on the points of the number and mass of the clumps (compare with $2 - 3~{\rm Gyr}$ in Fig.\ref{snap}). In our simulation, the mass resolution of gas, which corresponds to the minimum resolved Jeans mass, is approximately $N_{\rm nb}\times m_{\rm gas}\simeq2\times10^5~{\rm M_\odot}$. This is much smaller than the mean clump masses observed in \citet{ee:05} and comparable to the smallest clumps in \citet{eem:09}. Therefore, the mass and number of clumps we obtained in our simulation are do not suffer from the resolution effects.

However, current high-redshift observations still have difficulties of limited angular resolution and detectability, which would have caused underestimates of the number of clumps and the disc diameter. Moreover, low luminosity clump clusters might have been missed. As discussed in \citet{eem:09}, the current high-redshift observations would be biased to bright parts in bright galaxies. \citet{ee:05} discussed that their clump cluster samples have a too high inter-clump surface density to be a typical nearby spiral galaxy like the MW. Although clump clusters with a low surface brightness would not be the most representative cases for the high-redshift clump clusters, it may be possible that the low surface brightness clump clusters had been present but missed in the observations, and may be a progenitor of the nearby spiral galaxies. Our simulation result may corresponds to such a low luminosity clump cluster, which suggests the old and metal-rich pseudobulge formation from the clumps.

The longevity of the clumps in our simulation should be also examined carefully. In our simulation, the clumps take $\sim5~{\rm Gyr}$ to merge into the bulge. The clump migrations in the previous simulations have accomplished in a shorter time, $\sim 1~{\rm Gyr}$ \citep[e.g.][]{isg:04,bee:07,bem:09,ebe:08,abj:10,cdm:11}. The main causes of this difference are the disc size and the clamp mass. The clumps in our simulation have to migrate a long distance to the galactic centre because of the larger disc scale length than that in the previous works. In addition, the clump masses seem somewhat smaller than these in the previous works. Therefore, the dynamical friction on the clumps is weak in our simulation.

In such a case of the slow evolution with the long-lived clumps, the clump clusters can not finish the formation of disc and bulge in the early Universe. If the  evolution time scale is $\sim 5~{\rm Gyr}$ like our simulation, the clumps remain in galaxies until after the redshift $z\sim1$. Observationally, there exists clump clusters even after the redshift $z\sim1$, and \citet{eem:09} observed that the clump clusters in the lower-redshift Universe are less massive than these in the higher-redshift Universe although they discussed that this 'downsizing' is due to a detection limit in the observation and a smaller sampling volume in the lower-redshift Universe. If this downsizing is real, our simulation may correspond to such less massive galaxies evolving slowly until the low-redshift.

Disruption of the clumps should be also discussed. Recently, \citet{gnj:11} has observed strong outflows from some clumps, which are faster than the escape velocity of the clump. \citet{gng:10} demonstrated the clump disruption due to the superwinds implemented in their numerical simulations by a parametrised wind model, although \citet{kd:10} had analytically discussed that such a strong feedback would not be realistic. The SN feedback model adopted in our simulation can not reproduce such destructive superwinds. \citet{kd:10} suggested that the most probable engine of the superwinds is radiation pressure from newly formed massive stars. However, most of simulations for galaxy formation have not been implemented with the radiation pressure from the massive stars and our simulation did not either. The result of \citet{gng:10} implies that taking the radiative feedback into account may drastically change the picture of the simulations for clump clusters.

\subsection{Similarity to the MW bulge}
The MW bulge does not follow the properties of classical bulges nor pseudobulges. The MW bulge shows a nearly exponential profile, an oblate peanut shape (X-shape) and a significant rotation, which are similar to pseudobulges. At the same time, it has old stars, a high metallicity and weak star formation, which are similar to classical bulges \citep[e.g.][and references therein]{wgf:97,kk:04}. These properties of the MW bulge are consistent with the clump-origin bulge obtained by our simulation although the absence of cylindrical rotation is inconsistent with the MW bulge.

Bulge stars of the MW are observed to be $\alpha$-element enhanced (e.g. \citealt[][and references therein]{mr:94,kk:04}; \citealt{z:06,zhl:08,lhz:07,fmr:07,m:08,ama:10,h:11}). Interestingly, \citet{z:06} observationally discussed that an $\ofe$ vs. $\feh$ distribution of the MW bulge stars agrees very well with the result of the chemo-dynamical simulation for clump-origin bulge formation in \citet{isg:04}. On the other hand, contrary to \citet{z:06}, the recent observation of Baade's window by \citet{h:11} showed considerable disagreement in a metallicity distribution function with the simulation result of \citet{isg:04}. These chemical abundances will provide further constraints on the bulge formation scenario.

We assumed an idealized simulation model, a collapsing gas sphere in a Navarro-Frenk-White model halo, in order to study \textit{naive natures} of a clump-origin bulge. As we noted repeatedly in previous sections, our model would need the later gas accretion to form a thin disc \citep{bem:09}. Moreover, in the cosmological context, real galaxies suffer galaxy merger more or less. These physical processes following the clump-origin bulge formation could alter the naive natures of the clump-origin bulge. For example, as we noted in \S 3.1.3, a barred structure may induce a cylindrical rotation of the bulge \citep{smg:11}.

Recently, some observations showed bimodal metallicity distributions and kinematics among MW bulge stars, implying a multiple formation process of the MW bulge \citep{ba:10,b:10,b:11,h:11}. \citet{ba:10} and \citet{h:11} suggested that the MW bulge is a complex bulge of a classical bulge and a pseudobulge. Thus, bulge formation in real galaxies (including the MW) might be complex and proceed through multiple scenarios. Some bulges would be a composite structure as proposed by some observations of extra-galaxies \citep[e.g.][]{ebg:03,kdb:10,kb:10,fd:11}. 

Such unclassifiable bulges like the MW bulge (old pseudobulge) are also observed in some other disc galaxies \citep[e.g.][]{ba:87,pbd:99}. Our simulation results suggest that clump-origin bulges could be a possible origin of these old pseudobulges. In addition, as we discussed in \S 3.4, the inactive pseudobulges may be also clump-origin bulges.

\subsection{Exponential bulges in the high-redshift Universe}
It is very difficult to determine whether a bulge observed in a distant disc galaxy is clump-origin or not. Although \citet{kk:04} has argued that the secular evolution could build some pseudobulges quickly $\hspace{0.3em}\raisebox{0.4ex}{$>$}\hspace{-0.75em}\raisebox{-.7ex}{$\sim$}\hspace{0.3em}5~{\rm Gyr}$ ago, we expect that the secular evolution would not be able to build up a pseudobulge in a premature disc in early phase of the galaxy formation. It is  because the `\textit{secular}' evolution has to postdate accomplishment of their disc formation and takes a long time to form a pseudobulge. Therefore, we can expect that a galactic bulge observed in the high-redshift Universe would be either a classical bulge or a clump-origin bulge. As we showed in \S 3.1.1, the clump-origin bulge has a nearly exponential surface density profile, which is distinguishable from the $R^{1/4}$ law profile of the classical bulges. Thus, if the density profile of the bulges in clump clusters could be observationally resolved, we would be able to identify the clump-origin bulges by identifying their exponential density profile.

\citet{eer:07} has observationally indicated that comoving spatial number density of clumpy galaxies (clump clusters and chain galaxies) in the high-redshift Universe ($1\hspace{0.3em}\raisebox{0.4ex}{$<$}\hspace{-0.75em}\raisebox{-.7ex}{$\sim$}\hspace{0.3em}z\hspace{0.3em}\raisebox{0.4ex}{$<$}\hspace{-0.75em}\raisebox{-.7ex}{$\sim$}\hspace{0.3em}4$) are consistent with the density of spiral galaxies in the nearby Universe and suggested that quite a few galaxies of current spirals used to be a clump cluster or a chain galaxy. This implies that disc galaxies hosting a clump-origin bulge may not be rare. However, the idea that all disc galaxies must have formed a clump-origin bulge contradicts the observational fact that the number of bulge-less spiral galaxies is not small \citep{kdb:10,fd:11}. As we noted in \S4.1, \citet{ee:05} had discussed that the observed clump clusters could not be the nearby spiral galaxies because of the too high surface density. 

\subsection{Comparison with previous studies}
So far, there have been only a few studies discussing the natures of clump-origin bulges. As we noted in \S3.1, \citet{ebe:08} has argued with numerical simulations that clump-origin bulges have a similar dynamical state to classical bulges, such as a large S\'ersic index, $n\hspace{0.3em}\raisebox{0.4ex}{$>$}\hspace{-0.75em}\raisebox{-.7ex}{$\sim$}\hspace{0.3em}2$ and a slow rotation, $V/\sigma_0\simeq 0.4-0.5$. Their results indicating classical bulge-like signatures poses a discrepancy with our results. This disagreement may imply that the properties of clump-origin bulges depend on initial conditions of simulations. Our simulation model adopted the collapsing gas sphere model, while \citet{ebe:08} used a uniform density disc model as their initial condition. As we discussed, the size and number of the clumps would depend on the initial spin parameter. The size of the clumps may affect the relaxation process in the clump mergers. In the simulations of \citet{ebe:08}, fewer and larger clumps form in the clump cluster and the resultant $B/T$ ratio is somewhat larger than in our simulation.

Simulation method may also be a possible cause of the disagreement. The simulations of \citet{ebe:08} were performed by a particle-mesh sticky-particle code which mimics hydrodynamics by inelastic collisions between particles designated as a gas particle. On the other hand, we describe the multi-phase interstellar medium with our SPH code which treats radiative cooling, far-ultraviolet heating and SN feedback with hydrodynamical equations of motion. Since the sticky particle method is only an imitation of hydrodynamics, the evolution of the ISM represented by the sticky-particle method may lead different from that by the SPH method. The SPH simulations have also still suffered from the overcooling problem \citep[e.g.][]{kg:91,nb:91,nw:94,tw:95,tw:96,oef:05,sgs:11}. Degree of dissipation, i.e. richness of gas, in merging clumps may also affect the properties of the clump-origin bulge. Moreover, structure of remnants seems to depend on the equation of state and cooling of gas \citep{bo:11}. Hence, the contradiction of our results of the clump-origin bulge to that of \citet{ebe:08} may stem from a difference of gaseous physics in the clump merger. The degree of dissipation in the clump mergers may depend on initial conditions of a simulation model and/or the adopted numerical scheme for hydrodynamics (the sticky-particle or SPH code).

As noted above, the clump-origin bulge in our simulation has nearly twice larger rotation in terms of $V_{\rm max}/\sigma_0$ than the result of \citet{ebe:08}. Recently, the simulations of \citet{abj:10} and \citet{cdm:11} have also demonstrated that the clumps in clump clusters are dynamically rotation supported systems during the migration toward the galactic centre although they paid little attention to resultant bulges. Especially, \citet{cdm:11} mentioned that clumps which have experienced merger with other clumps tend to be highly supported by rotation and indicated that more massive clumps have larger rotation (their \S 4.2). Although \citet{cdm:11} discussed kinematics of gas only, their results imply that the clump merger can spin up the remnant clumps. Therefore, we could expect that the clump-origin bulges in their simulations would have a significant rotation because the clump-origin bulges are the final remnant of the clump mergers. 

Clump-origin bulges form through `\textit{mergers of stellar clumps}' in a disc, whereas classical bulges form through `\textit{mergers of galaxies}'. Formation of the clumps are caused by instability of gas in rotating disc, therefore, these clumps retain a memory of their discy origin and a spin in the same direction as the disc rotation as shown by \citet{cdm:11}. The bulge formation by the clump mergers takes place in the disc plane after orbiting around the galactic centre in the same direction, i.e. `\textit{prograde-prograde merger}'. As a result, the clump-origin bulges which are remnant of the clump merger retain the orbital angular momentum of the merger and the spin angular momentum of the progenitor clumps in their kinematics. Then, the bulges should acquire the significant rotation of which property is different from that of classical bulges although both of them are built through merger-origin formation scenarios. 

Our result indicated the nearly exponential surface density profile with the S\'ersic index of $n = 1.18$. We speculate that this exponential profile may stem from the elongated bar-like shape of the bulge in our simulation (Fig. \ref{bulge_zoom}). It has been observed that barred structures in spiral galaxies have an exponential density profile \citep[e.g.][]{ee:85,ohw:90}. This was demonstrated also using numerical simulations by \citet{ce:93} and \citet{n:96}\footnote{He proposed that spontaneous bars have an exponential profile, tidal bars have a flat one, in agreement with observations.}.


\section{Summary}
We summarize the natures of the clump-origin bulge in our simulation:
\begin{itemize}
\item nearly exponential surface density profile,
\item elongated bar-like structure, and boxy shape,
\item significant net rotation, but not cylindrical,
\item old stellar population,
\item high metallicity of stars with vertical gradient and 
\item inactive star formation (if there is no late accretion).
\end{itemize}
If we compare these natures with those observed in typical classical bulge and pseudobulges \citep{kk:04}, the former three natures, exponential profile, boxy shape and a significant rotation, indicate pseudobulge-like signatures. On the other hand, the latter three, old, metal-rich stars and weak star formation, represent classical bulge-like signatures. Therefore, the clump-origin bulge, which we obtained here, can not be simply classified into classical bulges nor pseudobulges, but should have unique properties as shown in this paper.

However, \citet{ebe:08} has suggested with numerical simulations that clump-origin bulges should be a classical bulge. This inconsistency would be due to adopting different initial conditions and/or numerical schemes. As we discussed in \S 3.1.1 and 4, initial setup in simulations would control disc diameters, gas density in the disc and the size of clump-origin bulges finally, and may affect the natures of the clump-origin bulge.

Hence, our results open up a different possibility of clump-origin bulges than the result of \citet{ebe:08}. In our simulation, the surface density of the disc is lower and the resultant bulge is smaller than \citet{ebe:08}. Therefore, it is expected that low surface brightness clump clusters prefer a \textit{clump-origin pseudobulge} and enable the galaxy to avoid the \textit{clump-origin classical bulge} formation. The clump-origin pseudobulge is more applicable to the MW bulge which is a pseudobulge-like but old metal-rich bulge. The low surface brightness clump clusters might had been present but missed in the observations and \citet{ee:05} discussed that clump clusters in their observation have a too high surface density to be the MW progenitor. Thus, to investigate the clump-origin bulge formation and its natures may take an important role to seek the formation history of the MW.

\section*{Acknowledgments}

We thank Fr\'ed\'eric Bournaud for his fascinating suggestions as the reviewer, who kindly helped improve the paper. We are grateful to Daisuke Kawata for polishing up the paper, Masafumi Noguchi for his helpful discussion. S.I. is financially supported by Research Fellowships of the Japan Society for the Promotion of Science (JSPS) for Young Scientists. The numerical simulations reported here were carried out on Cray XT-4 kindly made available by CfCA (Center for Computational Astrophysics) at the National Astronomical Observatory of Japan. This project is partly supported by HPCI Strategic Program Field 5 `The origin of matter and the universe'.


\end{document}